\documentclass[12pt]{iopart}

\usepackage{amsfonts}
\usepackage{amssymb}
\usepackage{epsfig}
\usepackage{graphicx}

\def\be{\begin{equation}}
\def\ee{\end{equation}}
\def\bea{\begin{eqnarray}}
\def\eea{\end{eqnarray}}
\def\ben{\begin{enumerate}}
\def\een{\end{enumerate}}
\def\bei{\begin{itemize}}
\def\eei{\end{itemize}}

\def\nn{\nonumber}
\def\mc{\mathcal}
\def\({\left(}
\def\){\right)}

\newcommand{\vo}{{\cal V}}

\begin{document}

\title[String moduli inflation: an overview]{String moduli inflation: an overview}

\author{Michele Cicoli${}^{1}$ and Fernando Quevedo${}^{2,3}$}

\address{$^1$ Deutsches Elektronen-Synchrotron DESY, Notkestrasse, 22607 Hamburg, Germany.}
\address{$^2$ DAMTP, University of Cambridge, Wilberforce Road, Cambridge CB3 0WA, UK.}
\address{$^3$ Abdus Salam ICTP, Strada Costiera 11, Trieste 34014, Italy.}
\ead{michele.cicoli@desy.de, F.Quevedo@damtp.cam.ac.uk}
\begin{abstract}
We present an overview of inflationary models derived from string theory
focusing mostly on closed string moduli as inflatons.
After a detailed discussion of the $\eta$-problem and different approaches to address it,
we describe possible ways to obtain a de Sitter vacuum with all closed string moduli stabilised.
We then look for inflationary directions and present some of the most
promising scenarios where the inflatons are either the real or
the imaginary part of K\"ahler moduli.
We pay particular attention on extracting potential observable implications,
showing how most of the scenarios predict negligible gravitational waves
and could therefore be ruled out by the Planck satellite.
We conclude by briefly mentioning some open challenges in string cosmology beyond
deriving just inflation.
\end{abstract}

\submitto{\CQG}

\maketitle

\section{Inflation and string theory}

There are several reasons for trying to derive inflationary models
from an ultra-violet complete theory such as string theory \cite{GloriousHistory}.
Some of the main motivations are:
\bei
\item Inflation generically involves energy scales which are much higher than those
of ordinary particle phenomenology, and so cosmology seems to be more promising
to directly probe string-related physics (unless the string scale is around the TeV
so that the LHC can put string theory to experimental test \cite{LHCstrings}).

\item The vast majority of inflationary models are based on the existence of
very shallow scalar potentials which give rise to extremely light scalar
masses, $m_{\phi} \ll H$. However, scalar masses are notoriously sensitive to
microscopic details since they generically get large contributions
when the short-distance sector of the theory is integrated out.
Due to this UV-sensitivity of inflation (often called `$\eta$-problem' \cite{etaproblem}),
model building can be trusted only via the knowledge of its UV completion.

\item String theory has many non-trivial constraints to model building,
in the sense that some signals seem to be generic in string-derived
effective Lagrangians, while others look very difficult to derive.
Therefore an observation of these unlikely signals would strongly constrain
inflationary model-building in string theory. Two main examples are
tensor modes and non-gaussianities since many, but \textit{not} all,
string inflationary models, predict a very small amount for these two
inflationary observables \cite{BMcA,SmallGW}.

\item The number of known field theoretic models of inflation could be
largely reduced and restricted by the requirement of sensible embedding
into string theory.

\item It is  reasonable to expect that inflation combined with particle
physics will help us to understand which vacuum of the string landscape
we live in, sheding also some light on the particular mechanism through which we end up there.

\item A complete understanding of reheating requires the knowledge
of all the relevant degrees of freedom at inflationary energies
in order to make sure that too much energy is not lost into hidden sector particles
\cite{warpedreheat,ClosedReheating}.

\item Given that the choice of special initial conditions is crucial
for most of the available inflationary models, the knowledge of the
full microscopic theory might help to address the origin of these initial conditions.

\item Particular inflationary models derived from string theory
predict purely stringy signals that cannot arise in any low-energy field theory.
A primary example is cosmic strings \cite{cosmicStrings} which
get formed at the end of brane/anti-brane inflation \cite{DDinflation,BBbarInfl}.
\eei

\section{Inflation and UV-sensitivity}

In this section we shall illustrate the origin of the $\eta$-problem in generic supergravity potentials
focusing on the simplest case of an inflaton which we take
as a real scalar field $\varphi$ ($\varphi^2=|\Phi|^2$ for a canonically normalised complex scalar $\Phi$)
in a 4D $\mc{N}=1$ supergravity. The inflationary potential
is given by the $F$-term scalar potential:
\be
V_F = e^{K/M_P^2}\left(K^{i\bar{j}}D_i W
D_{\bar{j}}\overline{W} -\frac{3|W|^2}{M_P^2}  \right).
\label{Vf}
\ee
Expanding the factor $e^K$ for $K = \varphi^2$ and
keeping the three leading order terms in (\ref{Vf}),
we can rewrite the inflationary potential as:
\be
V=\(1+\frac{\varphi^2}{M_P^2}+\frac{\varphi^4}{2 M_P^4}\)V_0(\varphi),
\label{Vexp}
\ee
and then compute its first and second derivatives:
\bea
V'&=&\(\frac{2\varphi}{M_P^2}+\frac{2\varphi^3}{M_P^4}\)V_0
+\(1+\frac{\varphi^2}{M_P^2}+\frac{\varphi^4}{2 M_P^4}\)V'_0, \nn \\
V''&=&\(\frac{2}{M_P^2}+\frac{6\varphi^2}{M_P^2}\)V_0
+2\(\frac{2\varphi}{M_P^2}+\frac{2\varphi^3}{M_P^4}\)V_0'
+\(1+\frac{\varphi^2}{M_P^2}+\frac{\varphi^4}{2 M_P^4}\)V''_0. \nn
\eea
We shall assume that an appropriate mechanism ensures the flatness of the potential $V_0(\varphi)$, so
that the corresponding slow-roll conditions are satisfied:
\be
\epsilon_0\equiv \frac{M_P^2}{2}\(\frac{V_0'}{V_0}\)^2\ll 1
\qquad \textrm{and} \qquad
\eta_0\equiv M_P^2\frac{V_0''}{V_0}\ll 1.
\ee
Defining $x\equiv \varphi/M_P$, we then obtain for the total slow-roll parameter $\eta$:
\be
\eta\equiv M_P^2\frac{V''}{V}=\eta_0 +\frac{4x(1+x^2)}{1+x^2+x^4/2}\sqrt{2\epsilon_0}+2\frac{1+3 x^2}{1+x^2+x^4/2}.
\label{etatotal}
\ee
Let us analyse this result separately for the two different kinds of inflationary models:
\bei
\item Small-field inflationary models: $\varphi\ll M_P$ $\Leftrightarrow$ $x\ll 1$. The expression (\ref{etatotal}) becomes:
\be
\eta\simeq \eta_0 + 4x\sqrt{2\epsilon_0}+2\simeq \mc{O}(1),
\label{etaSF}
\ee
and so the slow-roll conditions are violated.
In order to have $\eta\ll 1$ we should start with $\eta_0\sim -2 +\hat\eta$
with $\hat\eta\ll 1$ which generally requires fine-tuning of order $1\%$.

\item Large-field inflationary models: $\varphi\gg M_P$ $\Leftrightarrow$ $x\gg 1$.  Even though naively for this case the expression (\ref{etatotal}) becomes:
\be
\eta\simeq \eta_0 + \frac{8}{x}\sqrt{2\epsilon_0}+\frac{12}{x^2}\ll 1,
\label{etaLF}
\ee
higher order operators in the potential (\ref{Vexp}) become more and more important
and the series of  higher order operators will give rise to a contribution to the potential larger than $M_P^4$ even if $V_0 \ll M_P^4$.
In this case, generic string or Planck-scale corrections become significant.
This is just an indication that the series in $\varphi/M_P$ is not well defined
if $\varphi \gg M_P$ and the effective field theory (EFT) is not valid.
\eei
We stress that for small field inflationary models, the EFT is under control,
and so higher order operators are less and less important but they can still give rise
to large $\mc{O}(1)$ corrections to $\eta$ (see (\ref{etaSF})). On the other hand, for large field models,
the contribution to $\eta$ from dimension 6 and 8 operators are both still much less than unity (see (\ref{etaLF})),
but it is inconsistent to consider just these operators since higher dimensional ones will be even more important.
Thus the situation is even more complicated for large field inflation, since one has to check that higher order operators
are not generated not just to preserve inflation but also to have a consistent EFT.
Notice also that a canonical K\"ahler potential
naturally brings the potential to large values that invalidate the EFT independent of any expansion,
due to the exponential dependence of $V$ on $K$.

Hence we realised that inflation is even more UV-sensitive in the case of models with $\Delta\varphi\gg M_P$,
but why are we interested in this regime?
In order to get observable gravity waves from inflation, since Lyth derived the following bound \cite{Lyth}:
\be
\frac{\Delta\varphi}{M_P}\simeq \(\frac{r}{0.1}\)^{1/2},
\label{Lythbound}
\ee
which can be easily derived recalling the definition of the number of e-foldings $N_e$ and the ratio
between the amplitude of the tensor and scalar perturbations $r$:
\be
N_e=\frac{1}{M_P^2}\int\frac{V}{V'}\,d\varphi
\qquad \textrm{and} \qquad r=\frac{T}{S}=16\epsilon =8 M_P^2 \(\frac{V'}{V}\)^2,
\ee
which imply:
\be
\Delta N_e=\frac{1}{M_P^2}\frac{V}{V'}\Delta\varphi
\qquad \Rightarrow \qquad
\frac{\Delta\varphi}{M_P}=M_P\frac{V'}{V}\Delta N_e=\Delta N_e\sqrt{\frac{r}{8}}.
\label{Derive}
\ee
Substituting in (\ref{Derive}) the value $\Delta N_e\simeq 5$ for the scales relevant for the measurement of $r$,
we obtain the bound (\ref{Lythbound}).
Notice that $\Delta\varphi$ corresponds to $\Delta N_e\simeq 5$, and so $\Delta\varphi$ corresponding to $\Delta N_e\simeq 50 - 60$ is even bigger.
The present observational limit on $r$ is $r<0.2$ and the forecasts for future cosmological observations are $r\sim 0.1$ by the ESA satellite PLANCK,
$r\sim 0.01$ by the balloon experiments SPIDER, EPIC and BICEP, and $r\sim 0.001$ by the planned NASA satellite CMBPol \cite{Verde,rBounds}.

Hence we need a trans-planckian field range during inflation to see gravity waves. Notice that the value of $r$
fixes also the inflationary scale $M_{inf}$ since $M_{inf}\simeq M_{GUT} r^{1/4}$, and so an observation of $r$ would
correspond to a direct test of GUT-scale physics.

We conclude this section by stressing that in order to solve the $\eta$-problem we need two mechanisms: one to obtain a flat potential,
i.e. $\eta_0\ll 1$, and the other to protect the flatness of this potential against dangerous higher order operators.
If the $\eta$-problem is present, a $1\%$ fine-tuning will be needed.
This is not desirable but, if possible, is not much tuning.
A non-trivial question is if the fine-tuning is achievable and computable in string models.
A better output would be to find mechanisms in string theory that avoid the fine-tuning.

\section{The $\eta$-problem in string inflation}

String compactifications are characterised by the ubiquitous
presence of moduli which, after moduli stabilisation, emerge
as natural good candidates for inflaton fields.
Focusing on single-field slow roll inflation,
there are two broad classes of string inflationary models
according to the origin of the inflaton field \cite{Openclosedinflatons}:
\ben
\item Open string models \cite{DDinflation,BBbarInfl,DBI};

\item Closed string models \cite{KModInfl,kahler,Cicoli3}.
\een
In this section we shall review how the $\eta$-problem is solved in each class of models,
showing how closed string models are more promising than the ones
based on open moduli since, in the latter case, slow-roll can be achieved only via fine-tuning.

\subsection{Open string models}

The inflaton is a brane position modulus that parameterises the separation between two branes
or a brane and an anti-brane \cite{DDinflation, BBbarInfl,DBI}.
In this class of models it is hard to get a flat potential since $\eta_0\sim \mc{O}(1)$
but one can obtain $\eta_0\ll 1$ by means of warping.

However there is no symmetry that protects the inflaton potential from getting large contributions
from higher order operators. Let us see this in the illustrative example of
$D3$/$\overline{D3}$-brane inflation \cite{BBbarInfl}. The K\"ahler potential for the inflaton
$\varphi$, which gives the distance between the $D3$ and $\overline{D3}$-brane,
 and the volume modulus $T$ is:
\be
K=-3\ln\left[\(T+\bar{T}\)-\bar{\varphi}\varphi\right].
\label{Kopen}
\ee
The volume mode is fixed following the KKLT procedure \cite{kklt}: $T+\bar{T}=\langle\(T+\bar{T}\)\rangle$.
Therefore the K\"ahler potential (\ref{Kopen}) can be expanded as:
\be
K=-3\ln\left[\langle\(T+\bar{T}\)\rangle\left(1-\frac{\bar{\phi}\phi}{\langle\(T+\bar{T}\)\rangle}\right)\right]
=K_0+3\frac{\bar{\phi}\phi}{\langle\(T+\bar{T}\)\rangle},
\ee
where we defined $K_0\equiv -3\ln\langle\(T+\bar{T}\)\rangle$. The canonically normalised inflaton
field is $\phi_c=\frac{\sqrt{3}\varphi}{\sqrt{\langle\(T+\bar{T}\)\rangle}}$, and so also the $F$-term scalar
potential can be expanded as:
\be
V_F= e^{K_0} U(\varphi_c) e^{\bar{\varphi}_c\varphi_c}\simeq e^{K_0} U(\varphi_c)
\left(1+\frac{\bar{\varphi}_c\varphi_c}{M_P^2}\right),
\ee
with the last term in parenthesis giving rise to a large correction to $\eta$, $\delta\eta\sim\mc{O}(1)$.
Thus inflation can be achieved only by means of fine-tuning.
A great effort has been made in order to make sure
this fine-tuning is actually possible in string models \cite{baumann}.
This illustrates not only how difficult it is to obtain proper inflationary potentials
in this way, but also the level of sophistication of string theoretic models
which can address this point in explicitly calculations.

Moreover it is very hard to get large tensor modes due to bounds on field ranges.
Let us see the reason following a simple argument. Calling $x$ the radial position of the brane
and $L$ the characteristic size of the Calabi-Yau: $\textrm{Vol}_{CY}= L^6$, we have the
geometric bound $\Delta x<L$. The canonically normalised inflaton field is $\varphi=M_s^2 x$
implying that:
\be
\frac{\Delta \varphi}{M_P}=\Delta x\frac{M_s^2}{M_P}<\frac{L M_s^2}{M_P}.
\label{Dphi}
\ee
The string scale is related to the 4D Planck scale by dimensional reduction:
\be
M_s=\frac{M_P}{\sqrt{\textrm{Vol}_{CY} M_s^6}}=\frac{M_P}{L^3 M_s^3},
\qquad \Rightarrow \qquad
M_P=M_s^4 L^3.
\label{DimRed}
\ee
Substituting (\ref{DimRed}) in (\ref{Dphi}) we end up with:
\be
\frac{\Delta\varphi}{M_P}<\frac{L M_s^2}{L^3 M_s^4}=\(\frac{1}{L M_s}\)^2,
\ee
but in order to trust the EFT we need $L\gg l_s=M_s^{-1}$
which implies $L M_s\gg 1$, and so $\Delta\varphi\ll M_P$. Note that
this bound holds also for the warped case \cite{BMcA}.

\subsection{Closed string models}

The inflaton is either the real or the imaginary part of a closed string modulus
\cite{KModInfl,kahler,Cicoli3,Nflation,Grimm,SW,nemanja,cklq,OtherClosedStringInfl}.
The relevant 4D closed moduli of type IIB Calabi-Yau orientifold compactifications
with $D3/D7$-branes and $O3/O7$-planes are:
\bea
T_i&=&\tau_i+i\, b_i^+,\quad\tau_i=\textrm{Vol}(D_i),\quad b_i^+=\int_{D_i}C_4,\quad i=1,...,h_{1,1}^+, \nn \\
G_j&=&c_j-i Sb_j^-,\quad c_j=\int_{\hat{D}_j}C_2,\quad b_j^-=\int_{\hat{D}_j}B_2,\quad j=1,...,h_{1,1}^-, \nn
\eea
where $D_i$ denotes divisors of the internal three-fold, $C_2$ and $C_4$ are respectively the Ramond-Ramond 2- and 4-form,
and $B_2$ is the Neveu Schwarz-Neveu Schwarz 2-form. Notice that the Hodge numbers split under the orientifold action as $h_{1,1}=h_{1,1}^++h_{1,1}^-$,
and we neglected the axio-dilaton $S$ and the complex structure moduli $U$
since they acquire large masses $m_{S,U}\simeq m_{3/2}\gtrsim H$ via background fluxes \cite{gkp}.

\subsubsection{Real part of T-moduli}

Examples of inflaton candidates which are the real part of the K\"ahler moduli are
blow-up 4-cycles \cite{kahler}, fibration moduli \cite{Cicoli3} and the volume mode $\vo$ \cite{cklq}.
In this case it is possible to get $\eta_0\ll 1$ due to the no-scale structure of the the K\"ahler
potential if $h_{1,1}^+>1$ and keeping $\vo$ fixed during inflation, in the sense that
the inflaton is a combination
of the K\"ahler moduli orthogonal to the volume mode. In this way the presence of inflaton-dependent
higher order operators can be avoided.

Let us be more precise. The tree-level K\"ahler potential with the leading order $\alpha'$
correction reads \cite{bbhl}:
\be
K=K_0+\delta K_{(\alpha')}=-2\ln\(\vo+\frac{\xi}{2 g_s^{3/2}}\)
\simeq -2\ln\vo-\frac{\xi}{g_s^{3/2}\vo},
\ee
with $\xi\propto \frac{\(h_{1,2}-h_{1,1}\)}{(2\pi)^3}\sim \mc{O}(1)$,
while the tree-level flux-generated superpotential is just a constant
once the $S$ and $U$-moduli have been integrated out:
$W=W_0$. It is crucial to notice that $W_0$ does not depend on the $T$-moduli.
$K_0$ is of no-scale type, meaning that:
\be
\sum_{i,\bar{j}}K^{i\bar{j}}_0 K_{0,i} K_{0,\bar{j}}=3,
\ee
which in turn implies that at tree-level all the $\tau$-directions are flat since:
\bea
V&=&e^K\(\sum_{U,S}K^{\alpha\bar{\beta}}D_{\alpha}W D_{\bar{\beta}}\bar{W}
+\sum_{T}K^{i\bar{j}}_0 D_i W D_{\bar{j}}\bar{W}-3|W|^2\) \nonumber \\
&=& e^K \sum_{U,S}K^{\alpha\bar{\beta}}D_{\alpha}W D_{\bar{\beta}}\bar{W}\geq 0.
\eea
The $U$-moduli and the axio-dilaton $S$ are fixed supersymmetrically by imposing
$D_S W=D_U W=0$ and the remaining potential for the $T$-moduli vanishes.

The leading order $\alpha'$ correction $\delta K_{(\alpha')}\simeq -\frac{\xi}{g_s^{3/2}\vo}$
breaks the no-scale structure but lifting only the volume direction. This implies that $\(h_{1,1}^+-1\)$
directions are still flat, and so are natural inflaton candidates.

Moreover the tree-level K\"ahler potential $K_0=-2\ln\vo$ also depends only on a particular
combination of the moduli which is the volume mode. Thus it is possible to evade the
$\eta$-problem since no inflaton-dependent higher order operator gets generated from expanding the
prefactor $e^K$ of the $\mc{N}=1$ $F$-term scalar potential.

The next question to ask is if there are further perturbative corrections that might break the
no-scale structure. The answer is yes and they are string loop corrections to the K\"ahler potential
\cite{bhk,bhp,Cicoli1}:
\be
K=K_0+\delta K_{(\alpha')}+\delta K_{(g_s)}.
\ee
These corrections arise from 1-loop processes involving open strings stretching
from two different stacks of $D7/D3$-branes. These processes
can also be interpreted as the tree-level exchange of closed strings carrying Kaluza-Klein momentum.
Therefore the dependence on the K\"{a}hler moduli of these $g_s$ corrections
for an arbitrary Calabi-Yau has been conjectured to be $\delta K_{(g_s)}\sim m_{KK}^{-2}/\vo$
\cite{bhp} \footnote{Exact computations are possible only
for simple toroidal orientifolds \cite{bhk}. However this conjecture has passed several low-energy tests \cite{Cicoli1}.},
where $m_{KK}^{-2}$ gives the scaling of the 2-point function in string frame, while the
inverse power of the volume accounts for the Weyl rescaling from string to 4D Einstein frame.
The Kaluza-Klein scale can be written in terms of the K\"{a}hler moduli if we denote with $t$ the volume
of the 2-cycle transverse to the 4-cycle wrapped by the $D7$-brane: $m_{KK}\sim M_s/\ell \sim M_s/t^{1/2}$.

Thus we end up with a correction dangerously larger than the leading order $\alpha'$ one:
\be
\delta K_{(g_s)} \sim \sum_i \frac{t_i}{\vo}\,>\,\delta K_{(\alpha')} \sim \frac{1}{\vo},
\label{Kgs}
\ee
since we need each $t_i\gg 1$ in order to trust the EFT approach.
This implies that all $\tau$-direction might be lifted instead of just the volume one.

However this naive expectation turns out to be wrong since there is a generic cancellation in the
scalar potential of the leading order contribution of the $g_s$ corrections such that
$\delta V_{(\alpha')}\,>\,\delta V_{(g_s)}$ even if $\delta K_{(\alpha')}\,<\,\delta K_{(g_s)}$.
This cancellation, named `extended no-scale structure' \cite{Cicoli1}, is a property of the EFT
that is crucial to solve the $\eta$-problem.

Let us be a bit more precise about this cancellation. It can be shown that in the presence of
a generic correction to $K$, $K=K_0 + \delta K$, and a constant tree-level superpotential, $W=W_0$,
if $\delta K$ is a homogeneous function in the $t$-moduli of degree $n=-2$, then at leading order $\delta V=0$ \cite{Cicoli1}.
This statement can be proven by expanding the inverse K\"{a}hler metric $\mc{K}^{-1}=\left(\mc{K}_0 + \mc{\delta K}\right)^{-1}$
and then using the homogeneity ansatz. In such a way, the scalar potential can be expanded as:
\be
V = V_0 + V_1 + V_2 + ...,
\ee
where $V_0=0$ due to the well-known no-scale structure, whereas:
\be
V_1 = -\frac{1}{4}\frac{W_0^2}{\vo^2} n \left(n+2\right)\delta K,
\ee
which vanishes for $n=-2$ giving rise to the extended no-scale structure, and:
\be
V_2 =\frac{W_0^2}{\vo^2}\sum_i \frac{\partial^2 K_0}{\partial\tau_i^2}.
\ee
Notice that the $g_s$ correction (\ref{Kgs}) is indeed a homogeneous function of the $t$-moduli
of degree $n=-2$ since the volume scales as $\vo \sim t^3$. Moreover the first non-vanishing contribution
to the scalar potential scales as $V_2 \sim 1/\left(\vo^3 t \right)$, and so for $t\gg 1$
it is indeed subleading with respect to the $\alpha'$ contribution which behaves like $V_{(\alpha')}\sim 1/ \vo^3$.

Let us now illustrate these concepts in the simple case of just one modulus: $\vo = \tau^{3/2}= t^3$ with $t=\sqrt{\tau}$.
The superpotential is a constant $W=W_0$ while the K\"{a}hler potential including the leading $\alpha'$ and $g_s$
corrections looks like (with $\hat{\xi}\equiv \xi / g_s^{3/2}$):
\be
K= -2\ln\vo -\frac{\hat{\xi}}{\vo} +\frac{\sqrt{\tau}}{\vo}.
\ee
The corresponding $F$-term scalar potential becomes:
\be
V= 0 + \frac{\hat{\xi} W_0^2}{\vo^3} + 0 \cdot \frac{W_0^2 \sqrt{\tau}}{\vo^3} + \frac{W_0^2}{\sqrt{\tau}\vo^3} + \frac{W_0^2}{\tau^{3/2}\vo^3}.
\label{Vtotal}
\ee
The first vanishing contribution is due to the no-scale structure, the second term gives the leading $\alpha'$
correction, the second vanishing piece is the extended no-scale, the fourth term represents the
leading $g_s$ correction, while the last term gives the subleading string loop contribution.
It is now clear that $\delta V_{(\alpha')}\,>\,\delta V_{(g_s)}$ since $\delta V_{(g_s)}/\delta V_{(\alpha')}= \tau^{-1/2}\ll 1$.

The extended no-scale structure admits a low-energy interpretation in terms of the Coleman-Weinberg potential \cite{cw}:
\be
\delta V_{1-loop} \simeq 0 \cdot \Lambda^4 + \Lambda^2 \,\textrm{STr} \left(M^2\right) + \textrm{STr}\left(M^4\ln\left(\frac{M^2}{\Lambda^2}\right)\right),
\label{CW}
\ee
where the cut-off is given by the Kaluza-Klein scale:
\be
\Lambda=M_{KK} \simeq \frac{M_s}{\tau^{1/4}} \simeq \frac{M_P}{\vo^{2/3}},
\quad\textrm{and}\quad \textrm{STr}\left(M^2\right)\simeq m_{3/2}^2 \simeq M_P^2/\vo^2.
\ee
Notice that the first term in (\ref{CW}) is vanishing in any supersymmetric theory with the same number of bosonic
and fermionic degrees of freedom. The volume scaling of the Coleman-Weinberg potential therefore turns out to be:
\be
\delta V_{1-loop} \simeq 0 \cdot \frac{1}{\vo^{8/3}} + \frac{1}{\vo^{10/3}} + \frac{1}{\vo^4},
\ee
which matches exactly the scaling of the string loop corrections in (\ref{Vtotal}) once we make the
substitution $\tau = \vo^{2/3}$:
\be
\delta V_{(g_s)} \simeq 0 \cdot \frac{\sqrt{\tau}}{\vo^3} + \frac{1}{\sqrt{\tau}\vo^3} + \frac{1}{\tau^{3/2}\vo^3}
\simeq 0 \cdot \frac{1}{\vo^{8/3}} + \frac{1}{\vo^{10/3}} + \frac{1}{\vo^4}.
\ee
\newpage
From this very general analysis of all the possible corrections to the scalar potential, we learnt that
the inflationary potential can be generated in two ways:
\begin{enumerate}
\item via string loops like in the `Fibre inflation' model \cite{Cicoli3};

\item via tiny non-perturbative effects like in the `Blow-up inflation' model \cite{kahler}.
In this case the superpotential gets non-perturbative corrections of the form:
\be
W = W_0 + \sum_i A_i \,e^{-a_i T_i},
\ee
which can then be used to generate the inflationary potential but only if there are no loop corrections since they
tend to dominate the non-perturbative effects in the inflationary region: $\delta V_{(g_s)}\gg \delta V_{np}$.
A possible way-out to avoid the presence of open string loop corrections \footnote{Subleading closed string loop effects \cite{hg}
are likely to depend only on the volume mode.} would be to generate the inflationary potential via $ED3$
instantons instead of gaugino condensation on $D7$-branes, but one should then carefully check if a high enough inflationary
scale could also be obtained.
\end{enumerate}

\subsubsection{Imaginary part of T-moduli}

Examples of inflaton candidates which are the imaginary part of the K\"ahler moduli are the $C_4$-, $C_2$-
or $B_2$-axions. Except for the $B_2$-axions, their action enjoys a shift symmetry $a \to a + c$ which is
broken only non-perturbatively \cite{shiftInflation}. This implies that the tree-level K\"{a}hler potential does not depend on
the axion field $a$, providing a nice mechanism to forbid the appearance of higher order operators
from the expansion of the prefactor $e^K$ of the $F$-term scalar potential.

This is of course not enough to solve the $\eta$-problem since one has still to explain why $\eta_0\equiv M_P^2 V_0''/V_0 \ll 1$.
Here are some possible attempts:
\begin{enumerate}
\item[i)] Trying to use just one $C_4$-axion as the inflaton, it turns out that
$\eta_0\ll 1$ only for an axion decay constant larger than $M_P$, and so the potential is not flat enough.

\item[ii)] $\eta_0\ll 1$ can be obtained without requiring trans-planckian axion decay constants by
considering a racetrack superpotential like in the `Racetrack inflation' models \cite{KModInfl}.
Inflationary slow-roll can be achieved along a $C_4$-axion direction close to a saddle point
only by means of fine-tuning and no observable gravity waves get produced.

\item[iii)] $\eta_0\ll 1$ can also be obtained for sub-planckian axion decay constants by
using $N$ $C_4$-axions which combine together to give the inflaton direction like in the `N-flation' model \cite{Nflation}
where $r$ turns out to be large \footnote{See also \cite{Grimm} for a model with $N$ $C_2$-axions.}. However this model is very hard to realise since the real parts of the $T$-moduli have to be fixed
at an energy much larger than the axion potential. Moreover having a scalar potential (and not the superpotential)
as a sum of exponentials is a very strong assumption for supergravity actions.

\item[iv)] The $\eta$-problem can be solved with the help of monodromy by choosing a $C_2$-axion as the inflaton
like in the `Axion monodromy' model \cite{SW,nemanja}. The inflaton travels a trans-planckian distance in field space
leading to observable gravity waves.
\end{enumerate}

\newpage

\section{Inflationary model building for closed moduli}

We shall follow a two-step strategy in order to build promising string inflationary models:
\begin{itemize}
\item[A)] Find a de Sitter vacuum with all closed string moduli stabilised so that the effective
field theory and the whole scalar potential are under control;

\item[B)] Look for inflationary directions \footnote{For a general analysis of the supergravity conditions
for dS vacua see \cite{marta} while for inflation see \cite{martainf}.}.
\end{itemize}

\subsection{Moduli stabilisation}

The first historic example with all geometric
moduli fixed is the KKLT scenario \cite{kklt} where $K$ and $W$ take the form:
\be
K= K_0 = -2 \ln\vo,\qquad W=W_0+ \sum_i^{h_{1,1}} A_i \,e^{-a_i T_i},
\ee
and all K\"{a}hler moduli are stabilised non-perturbatively. Despite
all its fundamental achievements, this scenario exhibits some problems:
\begin{enumerate}
\item Need to fine-tune $W_0$ such that $W_0\sim W_{np} \sim e^{-a \tau}$. This fine-tuning guarantees
the presence of a minimum and the fact that one can safely neglect perturbative corrections to the K\"{a}hler
potential since:
\be
V_{np}\sim e^K W_0 W_{np},\,\,\,V_p\sim e^K W_0^2 K_p\,\,\Rightarrow\,\,\frac{V_{np}}{V_p}
\sim\frac{W_{np}}{W_0 K_p}\sim \frac{1}{K_p}\ll 1,
\ee
given that $K_p = \delta K_{(\alpha')}\sim \vo^{-1}\ll 1$ for $\vo\gg 1$.
However for natural values of $W_0\sim \mc{O}(1)\gg W_{np}$
this is not true anymore.

\item Need non-perturbative effects for each 4-cycle and this is very hard to get. In fact, $W_{np}$ gets
definitely generated only for rigid cycles, and in addition there is a tension between instanton
effects and the presence of chirality \cite{blumenhagen}.

\item Need to add $\overline{D3}$-branes to up-lift the initial AdS minimum to dS. However in this way
supersymmetry is broken explicitly with some debates about the possibility to work in a supersymmetric framework.
\end{enumerate}

A solution to all these problems is provided by the LARGE Volume Scenario (LVS) \cite{LVS} which is an extension
and a generalisation of the KKLT set-up. No fine-tuning of the background fluxes is necessary,
so $W_0\sim \mc{O}(1)$, and perturbative corrections to $K$ have to be taken into account.
Let us describe the simplest version of the LVS with $h_{1,1}=2$ for compactifications on a orientifold
of the Calabi-Yau $\mathbb{P}^4_{[1,1,1,6,9]}(18)$. The volume reads $\vo=\tau_b^{3/2}-\tau_s^{3/2}$.
The $\mc{N}=1$ $F$-term scalar potential
can be derived from:
\be
K=-2 \ln\left(\tau_b^{3/2}-\tau_s^{3/2}\right)-\frac{\xi}{g_s^{3/2}\vo},\,\,\,\,\,\,\,\,\,\,W=W_0+ A_s \,e^{-a_s T_s},
\ee
and after axion minimisation (which sets $\cos\left(a_s \langle b_s\rangle\right)=-1$) takes the form:
\be
V= \frac{\sqrt{\tau_s}a_s^2 A_s^2 e^{-2 a_s\tau_s}}{\vo}
-\frac{a_s A_s W_0 \tau_s e^{-a_s \tau_s}}{\vo^2}+\frac{\xi W_0^2}{g_s^{3/2}\vo^3}.
\label{LVSV1}
\ee
Notice that we need non-perturbative corrections only in $\tau_s$ which is a rigid blow-up 4-cycle \cite{Cicoli2}.
We can integrate out $\tau_s$ by imposing $\partial V /\partial \tau_s=0$ which gives $a_s \tau_s\simeq \ln\vo$.
Hence (\ref{LVSV1}) simplifies to:
\be
V = -\frac{W_0^2\left(\ln\vo\right)^{3/2}}{\vo^3}+\frac{\xi W_0^2}{g_s^{3/2}\vo^3}.
\ee
Now setting $\partial V /\partial \vo=0$ we obtain $\ln\vo\simeq \xi^{2/3}/g_s$ corresponding to an
exponentially large volume minimum localised at (for $g_s\simeq 0.1$):
\be
\tau_s \sim g_s^{-1} \sim \mc{O}(10),\qquad\vo\sim W_0 \,e^{a_s\tau_s}\sim W_0 \,e^{a_s/g_s}\gg 1.
\ee
Note that $a_s= 2\pi$ in the case of an $ED3$ instanton whereas $a_s=2\pi/N$ for gaugino condensation in
an $SU(N)$ theory. The minimum is AdS but, contrary to KKLT, it breaks SUSY spontaneously. The fact that SUSY is
broken can be realised by noticing that the scalar potential (\ref{LVSV1}) scales as $V\sim \mc{O}(\vo^{-3})$
while the general expression of the $F$-term scalar potential contains a terms that scales as
$-3 e^K |W|^2 \sim \mc{O}(\vo^{-2})$. This forces also some $D_i W$ to be non-zero so that a cancellation in $V$ at order
$\vo^{-2}$ can be obtained. A detailed calculation reveals that the largest $F$-term is the one of the volume mode,
$F_{T_b}\neq 0$, so that $\tau_b$ is the main responsible for SUSY breaking and its corresponding fermionic partner is the Goldstino
that gets eaten-up by the gravitino \cite{LVS2}.

The $D$-term scalar potential can be used to get a dS vacuum \cite{bkq,cremades}
\footnote{For possible scenarios with $F$-term uplifting see \cite{Fup}.}.
In fact wrapping a $D7$-brane around $\tau_b$ and turning
on a gauge flux on an internal 2-cycle, leads to the following $D$-term potential:
\be
V_D = g^2 \left(\xi_b - \sum_i q_i \frac{\partial K}{\partial \varphi_i} \varphi_i \right)^2,
\ee
where the Fayet-Iliopoulos term is given by (with $q_b$ denoting a gauge flux coefficient):
\be
\xi_b = q_b \frac{\partial K}{\partial \tau_b} = -\frac{3}{2} \frac{q_b}{\tau_b}.
\ee
It can be shown that $V_F$ forces all $\langle|\varphi_i|\rangle=0$, and so, recalling that $g^2=\tau_b/(4\pi)$,
the $D$-term scalar potential reduces to $V_D\simeq p/ \tau_b^3\simeq p/\vo^2$, which can be used as a nice up-lifting
potential by fine-tuning the coefficient $p\sim\mc{O}(\vo^{-1})$
so that $V_D$ also scales as $\vo^{-3}$. Notice that in supergravity a non-vanishing $D$-term can be obtained
only if the $F$-term is also non-zero. This does not happen in KKLT since
the original AdS minimum is supersymmetric, contrary to the LVS where the AdS minimum breaks SUSY.

We finally point out that in the limit $W_0\ll 1$ the LVS vacuum reduces to the KKLT one, so in this sense the LVS
can be seen as a generalisation of the KKLT scenario. The LVS can also be extended to more complicated topologies
allowing the presence of many rigid blow-up modes with non-perturbative effects \cite{LVS2}:
\be
\vo =\tau_b^{3/2} - \sum_j \lambda_j \tau_j^{3/2}.
\ee
Moreover, as we have already seen, string loop corrections are negligible
due to the extended no-scale structure.

\subsection{Four-cycle moduli as inflaton fields}

\subsubsection{Blow-up inflation}

The minimal field content for realising `Blow-up inflation' is $h_{1,1}=3$ \cite{kahler}. The volume is
$\vo=\alpha\left(\tau_1^{3/2}-\gamma_2\tau_2^{3/2}-\gamma_3\tau_3^{3/2}\right)$, and the K\"{a}hler and
superpotential read:
\be
K=-2\ln\left(\vo+\frac{\xi}{g_s^{3/2}\vo}\right),\qquad W=W_0 + A_2\, e^{-a_2 T_2} + A_3\, e^{-a_3 T_3}.
\ee
The resulting scalar potential looks like:
\be
V= \sum_{i=2}^3 a_i^2 A_i^2 \frac{\sqrt{\tau_i}}{\vo} e^{-2 a_i \tau_i}
-\sum_{i=2}^3 a_i A_i W_0 \frac{\tau_i}{\vo} e^{-a_i\tau_i} +\frac{\xi W_0^2}{g_s^{3/2}\vo^3},
\ee
and the minimum is located at:
\be
a_2\tau_2\simeq a_3\tau_3\simeq g_s^{-1},\qquad\vo\simeq W_0\sqrt{\tau_2}\,e^{a_2\tau_2}
\simeq W_0\sqrt{\tau_3}\,e^{a_3\tau_3}.
\ee
Displacing $\tau_2$ far from its minimum, this field can drive inflation
while $\tau_3$ keeps the volume minimum stable during the inflationary dynamics.
The potential in terms of the canonically normalised inflaton field
$\varphi/M_P \simeq \tau_2^{3/4}/\vo^{1/2}$ takes the form:
\be
V \simeq V_0 -\beta \left(\frac{\varphi}{M_P \vo}\right)^{4/3} e^{-a \vo^{2/3} \left(\frac{\varphi}{M_P}\right)^{4/3}}.
\ee
In order to get both $\epsilon\ll 1$ and $\eta\ll 1$ we need $\vo^{2/3}\varphi^{4/3}\gg M_P^{4/3}$ $\Leftrightarrow$
$\varphi \gg M_P / \vo^{1/2} = M_s$. This implies that we are dealing with a small-field inflationary model that
does not yield observably large gravity waves: $r\ll 1$. On the other hand, the spectral index is in the right region
$0.960<n_s<0.967$. It is interesting to notice that inflation takes place exactly in the region of field space where
the EFT can be trusted since:
\be
\varphi\gg\frac{M_P}{\vo^{1/2}}\,\Leftrightarrow\,\frac{\tau^{3/4}}{\vo^{1/2}}M_P\gg\frac{M_P}{\vo^{1/2}}
\,\Leftrightarrow\,\tau^{3/4}\gg 1.
\ee
The value of the Calabi-Yau volume is fixed by the requirement of generating enough density perturbations,
$\delta \rho/\rho \sim 10^{-5}$, and it turns out to be $\vo \sim \left(10^6 \div 10^7\right) \ell_s^6$.

A dangerous correction to this potential comes from string loops which behave as:
\be
\delta V_{(g_s)}\sim \frac{1}{\sqrt{\tau_2}\vo^3}\sim \frac{1}{\varphi^{2/3}\vo^{10/3}},
\ee
and lead to a correction to the $\eta$-parameter of the form (where $V_0 \sim \hat{\xi}/\vo^3$):
\be
\delta\eta \sim M_P^2 \frac{\delta V_{(g_s)}''}{V_0} \sim \frac{1}{\varphi^{8/3}\vo^{1/3}\hat{\xi}}.
\label{etaBUI}
\ee
This correction is larger than unity for small $\tau_2$ since substituting in (\ref{etaBUI}) the expression
for the canonically normalised inflaton, we end up with $\delta\eta \sim \vo/(\tau_2^2 \hat{\xi})\gg 1$,
for $\tau_2 \ll \sqrt{\vo}$.
This may be seen as a manifestation of the $\eta$-problem in the loop corrected potential.
If the brane configuration is such that these corrections are absent, then this mechanism works,
otherwise fine-tuning is needed in order to get slow-roll inflation.
Notice that $\delta\eta$ can become smaller than unity for larger values of $\tau_2$ but
then $\delta V_{(g_s)}\gg \delta V_{(np)}$ and the inflationary potential is completely different since it would be generated
by string loops instead of non-perturbative effects. Moreover, one should make sure not to hit the walls
of the K\"{a}hler cone by taking $\tau_2$ too large. We shall now consider a different model where the inflationary potential is
indeed generated by string loops.

\subsubsection{Fibre inflation}

The `Fibre inflation' model is realised for Calabi-Yaus which are K3 fibrations
over a $\mathbb{P}^1$ base with an additional blow-up mode, so that  $h_{1,1}=3$ \cite{Cicoli3}.
The volume is $\vo=\alpha\left(\sqrt{\tau_1}\tau_2-\gamma_3\tau_3^{3/2}\right)$ while the K\"{a}hler and
superpotential read:
\be
K=-2\ln\left(\vo+\frac{\xi}{g_s^{3/2}\vo}\right),\qquad W=W_0 + A_3\, e^{-a_3 T_3}.
\ee
The resulting scalar potential looks like:
\be
V= a_3^2 A_3^2 \frac{\sqrt{\tau_3}}{\vo} e^{-2 a_3 \tau_3}
-a_3 A_3 W_0 \frac{\tau_3}{\vo} e^{-a_3\tau_3} +\frac{\xi W_0^2}{g_s^{3/2}\vo^3},
\ee
and depends only on two moduli: $\tau_3$ and $\vo$, which can be stabilised at
$\tau_3 \sim 1/g_s$ and $\vo \sim W_0\sqrt{\tau_3}\,e^{a_3\tau_3}$. The direction
in the $(\tau_1,\tau_2)$-plane orthogonal to the volume is still flat, and so
it represents a perfect inflaton candidate. This direction can be lifted only
at subleading order by the inclusion of string loop corrections which take the form:
\be
\delta V_{(g_s)} =\left(\frac{A g_s^2}{\tau_1^2}-\frac{B}{\vo\sqrt{\tau_1}}
+\frac{C g_s^2\tau_1}{\vo^2}\right)\frac{W_0^2}{\vo^2},
\label{Vgs}
\ee
where $A$, $B$ and $C$ are unknown $\mc{O}(1)$ coefficients. The potential (\ref{Vgs})
fixes the K3 fibre modulus $\tau_1$ at $\tau_1 \sim g_s^{4/3} \,\vo^{2/3}$.
Working in the $(\tau_1,\tau_2)$-plane and displacing $\tau_1$ far from its minimum,
this field can drive inflation while $\tau_3$ keeps the volume minimum stable.
The inflationary potential in terms of the canonically normalised inflaton field
$\varphi/M_P \simeq \sqrt{3}\ln\tau_1/2$ looks like (with $R\sim g_s^4 \ll 1$ and $\beta\sim \mc{O}(1)$):
\be
V = \frac{\beta}{\vo^{10/3}} \left(3-4 \,e^{-\frac{\varphi}{\sqrt{3}M_P}} +e^{-\frac{4\varphi}{\sqrt{3}M_P}}
+R \,e^{\frac{2\varphi}{\sqrt{3}M_P}}\right).
\label{VFI}
\ee
The inflationary dynamics is better understood dividing the field space in four different regions as indicated
in Figure 1 and then analysing each region separately.
\begin{figure}[ht]
\begin{center}
\epsfig{file=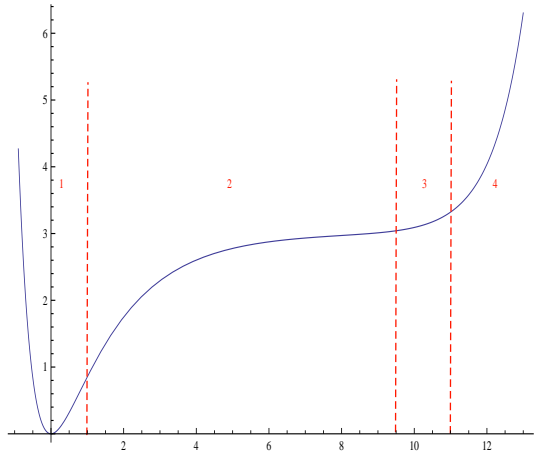, height=60mm,width=70mm}
\label{Loops}
\end{center}
\caption{Inflationary potential for the `Fibre inflation' model.}
\end{figure}
Inflation takes place in region 2 where the dynamics is completely dominated by the first negative exponential in (\ref{VFI}),
and so the inflationary potential can very well be approximated as:
\be
V_{inf}\simeq \frac{\beta}{\vo^{10/3}}\left(3-4\, e^{-\frac{\varphi}{\sqrt{3}M_P}}\right).
\ee
The boarder between region 1 and 2 is given by an inflectionary point with $\eta=0$ which
develops when the second negative exponential in (\ref{VFI}) starts competing with the first one.
The parameter $\epsilon$ becomes of the order unity and so inflation ends.
On the other hand, region 3 is characterised by the fact that the positive exponential in (\ref{VFI})
starts becoming important leading to $\eta\gg \epsilon >0$. We are still in the slow-roll regime
but the spectral index $n_s=1+2\eta-6\epsilon$ becomes larger than unity in contradiction with
CMB observations. Hence we do not consider any inflationary dynamics in this region. Finally
the boarder between region 3 and 4 is set by the violation of the slow-roll conditions.
Further away in this final region fully dominated by the positive exponential,
perturbation theory breaks down due to the fact that the K3 fibre modulus $\tau_1$ becomes
extremely large while the 2-cycle which is the basis of the fibration shrinks to zero size.

This is a nice example of large field inflation since $\Delta \varphi\gg M_P$.
Furthermore, all adjustable parameters enter only in the prefactor of the inflationary potential,
making this model very predictive and leading to an interesting relation between $r$ and $n_s$:
\be
r\simeq 6\left(n_s-1\right)^2.
\ee
The prediction for the cosmological observables is $n_s\simeq 0.97$ and $r\simeq 0.005$.
The inflationary scale is fixed by the requirement of generating the observed density perturbations,
and sets $\vo \sim 10^4 \ell_s^6$ corresponding to $M_s\simeq M_{GUT} \sim 10^{16}$ GeV.

Contrary to `Blow-up inflation', this mechanism does not suffer from the $\eta$-problem
in the sense that no $\mc{O}(1)$ corrections to $\eta$ arise at loop level.
Being a large field model, loop corrections are suppressed by the size of the field
and are naturally small. The $\eta$-problem is absent because
the supergravity potential being almost no-scale is {\it not} generic,
while the arguments used regarding the $\eta$-problem assume a generic K\"ahler potential.
There is also a symmetry enhancement in the limit of infinite volume.

\subsection{Axions as inflaton fields}

As we have already pointed out, the action of an axion field $a$
enjoys a shift symmetry $a \to a + c$ which is perturbatively exact. Hence no axion-dependent
dimension 6 $M_P$-suppressed operator of the form $V a^2 /M_P^2$ can arise, singling out the
axion as a natural inflaton candidate in relation to this nice partial solution of the $\eta$-problem.

\subsubsection{Single axion}

A typical axion potential is given by:
\be
V(a) = \Lambda^4\left(1-\cos\left(\frac{a}{f_a}\right)\right),
\label{AxionV}
\ee
where $f_a$ denotes the axion decay constant which sets its coupling to gauge bosons:
\be
\mc{L}\simeq \frac{1}{2}\,\partial_{\mu}a\,\partial^{\mu}a + \frac{a}{f_a} \int F\wedge F.
\ee
In the case of $C_4$-axions $b_i = \int_{\Sigma_i} C_4$, where $\Sigma_i$ is an internal 4-cycle,
the axion potential might be generated by a stringy instanton wrapping $\Sigma_i$. As we have already seen,
the superpotential for the LVS is $W= W_0+ A_s \,e^{-a_s T_s}$ with $T_s=\tau_s+ i b_s$,
and so the scalar potential in units of $M_P$ becomes (ignoring $\alpha'$ corrections):
\be
V= \frac{a_s^2 A_s^2 \sqrt{\tau_s} \,e^{-2 a_s \tau_s}}{\vo} - a_s A_s W_0 \frac{\tau_s
\,e^{-a_s\tau_s}}{\vo^2}\cos\left(a_s b_s\right).
\ee
We can therefore reproduce a potential of the form (\ref{AxionV}) with the scale $\Lambda$
given by:
\be
\Lambda^4 \sim \frac{e^{-a_s\tau_s}M_P^4}{\vo^2} \sim \frac{M_P^4}{\vo^3}\,\Rightarrow\,\Lambda \sim \frac{M_P}{\vo^{3/4}}.
\ee
The slow-roll parameters for the axion potential (\ref{AxionV}) look like:
\be
\epsilon= \frac{M_P^2}{2 f_a^2}\frac{\sin^2\left(a/f_a\right)}{\left(1-\cos\left(a/f_a\right)\right)^2}
\simeq \left(\frac{M_P}{f_a}\right)^2,
\ee
\be
\eta= \frac{M_P^2}{f_a^2}\frac{\cos\left(a/f_a\right)}{1-\cos\left(a/f_a\right)}
\simeq \left(\frac{M_P}{f_a}\right)^2.
\ee
showing how $\epsilon\simeq \eta \simeq \left(M_P/f_a\right)^2\ll 1$ only if $f_a\gg M_P$, but this is never the case.
Thus we realise that the potential of a single axion is not flat enough to drive inflation.

\subsubsection{Racetrack inflation}

The `Racetrack inflation' models were the first explicit examples of closed string inflation \cite{KModInfl}.
They were both realised in the context of the KKLT scenario with one and two K\"ahler
moduli respectively. In the first scenario the single-field
KKLT framework is modified by including a racetrack superpotential \cite{geq}:
\be
W=W_0 + A_1\, e^{-a_1 T} + A_2\, e^{-a_2 T}.
\ee
The scalar potential has a saddle point in the direction of the axion component of $T$.
By a tuning of the parameters of order $0.1\%$, the region close to the saddle point
can be made flat enough to allow (topological eternal \cite{topol}) inflation.
The main observational implications are a spectral index $n_s\simeq 0.95$
and negligible tensor modes ($r\leq 10^{-12}$).

Very similar physical implications are obtained in the improved `Racetrack inflation' scenario.
In this case the Calabi-Yau manifold has two K\"ahler moduli
$T_1$ and $T_2$ with K\"ahler and superpotential of the form:
\be
K=-2\ln\left(\tau_1^{3/2}-\tau_2^{3/2} \right),\qquad W=W_0 + A_1 \,e^{-a_1 T_1} + A_2\, e^{-a_2 T_2}.
\ee
In this case there are two axion-like fields and there is much less room to play with the parameters.
However inflation is also obtained in one of the axionic directions. The amount of fine-tuning is similar
to the single field case as well as all the physical implications.
In particular the value of the spectral index fits perfectly
well within the experimentally allowed window.

Both these scenarios were important
to realise inflation in a concrete string model even if
the amount of fine-tuning of the underlying parameters is significant.

\subsubsection{N-flation}

The first inflationary scenario based on more than one axion, say for definiteness $N$ of them,
is the `N-flation' model \cite{Nflation}. The total action takes the form:
\be
\mc{L}\simeq \sum_{i=1}^N \left[\frac{1}{2}\left(\partial a_i\right)^2
-\Lambda^4\left(1-\cos\left(\frac{a_i}{f_a}\right)\right)\right].
\ee
The equations of motion $\ddot{a_i}+3 H a_i = - \partial V / \partial a_i$ imply that each axion evolves independently.
However focusing on the collective motion of the $N$ axions we obtain:
\be
\epsilon\simeq \left(\frac{M_P}{f_a}\right)^2\frac{1}{N^2},\qquad\eta=\left(\frac{M_P}{f_a}\right)\frac{1}{N},
\ee
which gives rise to the possibility to get $\epsilon\ll 1 $ and $\eta\ll 1$ for $f_a \leq M_P$ if $N\gg 1$.
Moreover one can get large tensor modes even if each single axion travels a sub-planckian distance in field
space: $\Delta \varphi_a \ll M_P$.

The main obstacle against the realisation of such a multiple axion inflationary scenario is the need to fix
all the real parts of the $T$-moduli at an energy larger than the axion potential: $V(\tau)\gg V(a)$ $\Leftrightarrow$
$m_{\tau}\gg m_a$. This is the only way to avoid the presence of dangerous steep directions along which the field would
quickly roll down destroying the inflationary dynamics. However in general the non-perturbative effects used to
generate the axion potential give also a mass to the 4-cycle moduli of the same order of magnitude: $m_{\tau}\sim m_a$.
For example focusing again on the LVS, $m_{\tau_s}\sim m_{b_s}\sim M_P/\vo\sim m_{3/2}$.
Thus one should try to fix all the $\tau$'s perturbatively, which is a hard task without invoking fine-tuning.
Another issue is that one should check that these axions do not get eaten-up by any Green-Schwarz mechanism.

\subsubsection{Axion monodromy}

The first example of a single axion with a flat potential and $\Delta\varphi_a \gg M_P$ has
been provided by the `Axion monodromy' model \cite{SW}.
The key-ingredient to obtain a trans-planckian motion for a field like an axion
which is periodic with period given by its decay constant: $\varphi_a\to\varphi_a+f_a$,
is to use monodromy, i.e. studying how the axion behaves when it turns around a singularity.

In the case of $D5$-branes wrapping a 2-cycle $\Sigma_2$ of size ${\rm Vol}\left(\Sigma_2\right)=L_2^2$,
the action for the axion $b=\int_{\Sigma_2}B_2$ is:
\be
S^{D5}_{DBI}\sim -\frac{1}{g_s}\int_{M_4\times \Sigma_2} d^6\xi \sqrt{{\rm det}\left(G+B\right)}=
-\frac{1}{g_s}\int d^4 x \sqrt{-g}\sqrt{L_2^4 + b^2}.
\ee
At large $b$, the potential can be approximated as $V\sim \mu^3 b f = \mu^3 \varphi$,
becoming linear in the canonically normalised inflaton $\varphi$.
The $D5$-brane breaks the shift symmetry and gives a non-periodic contribution as
$b\to b+ f$. Thus $V$ undergoes a monodromy that `unwraps' the axion circle.
However $b$ appears also in the tree-level K\"{a}hler potential $K_0$,
and so it is not a good inflaton candidate
due to the presence of higher order operators.

A better candidate is the $C_2$-axion $c=\int_{\Sigma_2}C_2$ since it does not appear
in $K_0$. Wrapping an $NS5$-brane around $\Sigma_2$, the potential for the axion $c$ looks like:
\be
S_{NS5}\sim-\frac{1}{g_s}\int d^4 x \sqrt{-g}\sqrt{L_2^4+ g_s^2 c^2}.
\ee
Again at large $c$, the potential can be approximated as $V(c)\sim c$,
becoming linear in the canonically normalised inflaton.
The inclusion of non-perturbative effects in the K\"{a}hler potential from an $ED1$ instanton
wrapping $\Sigma_2$ gives rise to a cosine modulation:
\be
V\sim \mu^3 \varphi + \Lambda^4 \cos\left(\frac{\varphi}{f}\right).
\ee
One needs again to fix the real parts of the $T$-moduli at a higher scale
to avoid a possible destabilisation of the inflationary potential, which can
be done using warping. The spectral index is $n_s\simeq 0.975$
and, due to a trans-planckian field range, $\Delta\varphi\simeq 11 M_P$,
observable tensor modes with $r\simeq 0.07$ get produced. Moreover the
cosine modulation gives ripples in the power-spectrum \cite{McW}
and resonant non-gaussianities \cite{NGinAM}.
However this very promising inflationary model is missing a fully
consistent compact example.

\section{Open issues beyond inflation}

There are several open challenges in string cosmology beyond
deriving just closed moduli inflation.
In this section we shall just briefly mention some of them:
\begin{itemize}
\item Study of pre- and re-heating which is at the moment available only for brane-antibrane inflation \cite{warpedreheat}
and K\"{a}hler moduli inflation \cite{ClosedReheating}.

\item Study of finite-temperature corrections to the moduli potential from the thermal bath.
In this way one finds a maximal temperature $T_{max}$
that sets an upper bound on the reheat temperature, $T_{RH}<T_{max}$,
to avoid a decompactification limit. In the KKLT case $T_{max}\sim \sqrt{m_{3/2}M_P}\sim 10^{10}$ GeV
for $m_{3/2}\sim 1$ TeV \cite{BHLR},
while for the LVS case, $T_{max}\sim \left(m_{3/2}^3 M_P\right)^{1/4}\sim 10^7$ GeV \cite{Cicoli4}.
It would also be interesting to find a model where finite-temperature corrections generate a
late epoch of thermal inflation \cite{TI} which might be needed to eliminate
relics in the later Universe.

\item Study of the post-inflationary (non)-thermal history of the moduli.
It is important to follow the moduli evolution in the presence of thermal corrections \cite{barreiro},
and then study their late time decay with the possible production of nice astrophysical signals \cite{CQ}
or non-thermal dark matter \cite{Acharya}.

\item Derivation of inflationary scenarios which allow TeV-scale supersymmetry
solving the generic tension between cosmology and particle phenomenology \cite{linde}.
The origin of this tension can be illustrated expressing the inflationary scale
as $M_{inf}=V^{1/4}\simeq \sqrt{H M_P}$ and noticing that the scalar potential
in the KKLT and LVS cases scale as:
\be
V_{KKLT}\simeq m_{3/2}^2 M_P^2,\qquad V_{LVS}\simeq m_{3/2}^3 M_P.
\ee
Thus for KKLT: $m_{3/2}\sim H \sim M_{inf}^2 / M_P$ while for LVS: $m_{3/2}\sim H^{2/3} M_P^{1/3} \sim M_{inf}^{4/3}/ M_P^{1/3}$.
The requirement of generating the density perturbations, $\delta\rho/\rho\sim 10^{-5}$,
generically sets $M_{inf}\sim M_{GUT}$, which implies $M_{soft}\sim m_{3/2}\gg 1$ TeV.
\newpage
Two available solutions are:
\begin{enumerate}
\item The volume $\vo$ can be used as the inflaton field \cite{cklq}.
In this way a high inflationary scale can be obtained by having $\vo$ small during inflation,
while a low $m_{3/2}$ can be achieved for $\vo$ large at the end of inflation.
Unfortunately this scenario requires fine-tuning.

\item Some MSSM realisations via $D$-branes at orbifold singularities can be sequestered
from the bulk, resulting in a hierarchy between $M_{soft}$ and $m_{3/2}$ which might
be even of the order $M_{soft}\sim m_{3/2}/\vo$ \cite{sequestering}.
These models allow $M_{soft}\sim 1$ TeV for large $m_{3/2}$.
\end{enumerate}

\item Study of multi-field inflationary dynamics that can provide
alternative mechanisms \cite{curvaton,ModReheat} to generate the density perturbations,
which can lower the inflationary scale and give rise to large non-Gaussianities.
Two interesting scenarios are:

\begin{enumerate}
\item The authors of \cite{StringyCurvaton} derived a curvaton model \cite{curvaton}
from string theory by combining some features of `Blow-up inflation'
with others of `Fibre inflation'. The model predicts large non-gaussianities of local type,
but the inflationary scale is still of the order $M_{GUT}$.

\item A very promising string-inspired scenario is based on modulated reheating \cite{ModReheat}
but no explicit string derivation is presently available even if some attempts are currently
under way \cite{StringyMR}.
\end{enumerate}
\end{itemize}


\section*{References}
\begin{thebibliography}{10}

\bibitem{GloriousHistory}
  F.~Quevedo,
%  ``Lectures on string/brane cosmology,''
  Class.\ Quant.\ Grav.\  {\bf 19} (2002) 5721
  [arXiv:hep-th/0210292];
  %%CITATION = CQGRD,19,5721;%%
R.~Kallosh,
%  ``On Inflation in String Theory,''
  Lect.\ Notes Phys.\  {\bf 738} (2008) 119
  [arXiv:hep-th/0702059];
  %%CITATION = LNPHA,738,119;%%
C.~P.~Burgess,
%  ``Lectures on Cosmic Inflation and its Potential Stringy Realizations,''
  PoS {\bf P2GC} (2006) 008
  [Class.\ Quant.\ Grav.\  {\bf 24} (2007) S795]
  [arXiv:0708.2865 [hep-th]];
  %%CITATION = CQGRD,24,S795;%%
L.~McAllister and E.~Silverstein,
%  ``String Cosmology: A Review,''
  Gen.\ Rel.\ Grav.\  {\bf 40} (2008) 565
  [arXiv:0710.2951 [hep-th]];
  %%CITATION = GRGVA,40,565;%%
M.~Cicoli,
%``String Loop Moduli Stabilisation and Cosmology in IIB Flux Compactifications,''
Fortsch.\ Phys.\  {\bf 58} (2010) 115
  [arXiv:0907.0665 [hep-th]].
  %%CITATION = FPYKA,58,115;%%

\bibitem{LHCstrings}
D.~Lust \etal,
%S.~Stieberger and T.~R.~Taylor,
%  ``The LHC String Hunter's Companion,''
  Nucl.\ Phys.\  B {\bf 808} (2009) 1
  [arXiv:0807.3333 [hep-th]];
  %%CITATION = NUPHA,B808,1;%%
D.~Lust \etal,
%O.~Schlotterer, S.~Stieberger and T.~R.~Taylor,
%  ``The LHC String Hunter's Companion (II): Five-Particle Amplitudes and Universal Properties,''
  Nucl.\ Phys.\  B {\bf 828} (2010) 139
  [arXiv:0908.0409 [hep-th]];
  %%CITATION = NUPHA,B828,139;%%
M.~Cicoli \etal,
%C.~P.~Burgess and F.~Quevedo,
  %``Anisotropic Modulus Stabilisation: Strings at LHC Scales with Micron-sized
  %Extra Dimensions,''
  arXiv:1105.2107 [hep-th].
  %%CITATION = ARXIV:1105.2107;%%

\bibitem{etaproblem}
  E.~J.~Copeland \etal,
%, A.~R.~Liddle, D.~H.~Lyth, E.~D.~Stewart and D.~Wands,
%  ``False vacuum inflation with Einstein gravity,''
  Phys.\ Rev.\  D {\bf 49} (1994) 6410 [arXiv:astro-ph/9401011].
  %%CITATION = PHRVA,D49,6410;%%

\bibitem{BMcA}
  D.~Baumann and L.~McAllister,
 % ``A microscopic limit on gravitational waves from D-brane inflation,''
  Phys.\ Rev.\  D {\bf 75} (2007) 123508
  [arXiv:hep-th/0610285].
  %%CITATION = PHRVA,D75,123508;%%

\bibitem{SmallGW}
  R.~Kallosh and A.~Linde,
 % ``Testing String Theory with CMB,''
  JCAP {\bf 0704} (2007) 017
  [arXiv:0704.0647 [hep-th]].
  %%CITATION = JCAPA,0704,017;%

\bibitem{warpedreheat}
N.~Barnaby \etal,
%, C.~P.~Burgess and J.~M.~Cline,
%  ``Warped reheating in brane-antibrane inflation,''
  JCAP {\bf 0504} (2005) 007 [arXiv:hep-th/0412040];
  %%CITATION = JCAPA,0504,007;%%
L.~Kofman and P.~Yi,
%  ``Reheating the universe after string theory inflation,''
  Phys.\ Rev.\  D {\bf 72} (2005) 106001
  [arXiv:hep-th/0507257];
  %%CITATION = PHRVA,D72,106001;%%
D.~Chialva \etal,
%, G.~Shiu and B.~Underwood,
%  ``Warped reheating in multi-throat brane inflation,''
  JHEP {\bf 0601} (2006) 014
  [arXiv:hep-th/0508229].
  %%CITATION = JHEPA,0601,014;%%

\bibitem{ClosedReheating}
 D.~R.~Green,
  %``Reheating Closed String Inflation,''
  Phys.\ Rev.\  D {\bf 76} (2007) 103504
  [arXiv:0707.3832 [hep-th]];
  %%CITATION = PHRVA,D76,103504;%%
N.~Barnaby \etal,
 % ``Preheating After Modular Inflation,''
  JCAP {\bf 0912} (2009) 021
  [arXiv:0909.0503 [hep-th]];
  %%CITATION = JCAPA,0912,021;%%
M.~Cicoli and A.~Mazumdar,
 % ``Reheating for Closed String Inflation,''
  JCAP {\bf 1009} (2010) 025 [arXiv:1005.5076 [hep-th]];
  %%CITATION = JCAPA,1009,025;%%
  M.~Cicoli and A.~Mazumdar
%  ``Inflation in string theory: A Graceful exit to the real world,''
  Phys.\ Rev.\  D {\bf 83} (2011) 063527
  [arXiv:1010.0941 [hep-th]].
  %%CITATION = PHRVA,D83,063527;%%

\bibitem{cosmicStrings}
  M.~B.~Hindmarsh and T.~W.~B.~Kibble,
 % ``Cosmic strings,''
  Rept.\ Prog.\ Phys.\  {\bf 58} (1995) 477
  [arXiv:hep-ph/9411342].
  %%CITATION = RPPHA,58,477;%%

\bibitem{DDinflation}
G.~R.~Dvali and S.~H.~H.~Tye,
%  ``Brane inflation,''
  Phys.\ Lett.\  B {\bf 450} (1999) 72 [arXiv:hep-ph/9812483].
  %%CITATION = PHLTA,B450,72;%%

\bibitem{BBbarInfl}
C.~P.~Burgess \etal,
%  ``The inflationary brane-antibrane universe,''
  JHEP {\bf 0107} (2001) 047 [arXiv:hep-th/0105204];
  %%CITATION = JHEPA,0107,047;%%
G.~R.~Dvali \etal,
%, Q.~Shafi and S.~Solganik,
%  ``D-brane inflation,''
arXiv:hep-th/0105203;
  %%CITATION = HEP-TH/0105203;%%
C.~Herdeiro \etal,
%, S.~Hirano and R.~Kallosh,
%  ``String theory and hybrid inflation / acceleration,''
  JHEP {\bf 0112} (2001) 027 [arXiv:hep-th/0110271];
  %%CITATION = JHEPA,0112,027;%%
C.~P. Burgess \etal,
%``Brane antibrane inflation in orbifold and orientifold models,''
JHEP {\bf 0203} (2002) 052 [arXiv:hep-th/0111025];
%%CITATION = HEP-TH 0111025;%%
S.~Kachru \etal,
%``Towards inflation in string theory,''
JCAP {\bf 0310} (2003) 013 [arXiv:hep-th/0308055];
%%CITATION = HEP-TH 0308055;%%.
H.~Firouzjahi and S.~H.~H. Tye,
%``Closer towards inflation in string theory,''
 Phys.\ Lett.\  B {\bf 584} (2004) 147 [arXiv:hep-th/0312020];
  %%CITATION = PHLTA,B584,147;%%
N.~Iizuka and S.~P.~Trivedi,
  %``An Inflationary model in string theory,''
  Phys.\ Rev.\  D {\bf 70} (2004) 043519 [arXiv:hep-th/0403203];
  %%CITATION = PHRVA,D70,043519;%%
C.~P. Burgess \etal,
%, J.~M. Cline, H.~Stoica, and F.~Quevedo,
%``Inflation in realistic D-brane models,''
JHEP {\bf 09} (2004) 033 [arXiv:hep-th/0403119].

\bibitem{Lyth}
 D.H. Lyth,
 Phys. Rev. Lett. {\bf 78} (1997) 1861 [arXiv:hep-ph/9606387].
 %%CITATION = HEP-PH/9606387;%%

\bibitem{Verde}
  L.~Verde, H.~Peiris and R.~Jimenez,
%  ``Optimizing CMB polarization experiments to constrain inflationary physics,''
  JCAP {\bf 0601} (2006) 019
  [arXiv:astro-ph/0506036].
%; See also fig. 2 of \verb"www.b-pol.org/pdf/BPOL_Proposal.pdf".
  %%CITATION = JCAPA,0601,019;%%

\bibitem{rBounds}
J.~Bock \etal,
%  ``The Experimental Probe of Inflationary Cosmology (EPIC): A Mission Concept
%  Study for NASA's Einstein Inflation Probe,''
arXiv:0805.4207 [astro-ph].
  %%CITATION = ARXIV:0805.4207;%%

\bibitem{Openclosedinflatons}
P.~Binetruy and M.~K.~Gaillard,
  %``Candidates for the Inflaton Field in Superstring Models,''
  Phys.\ Rev.\  D {\bf 34}, 3069 (1986);
  %%CITATION = PHRVA,D34,3069;%%
 T.~Banks \etal,
%, M.~Berkooz, S.~H.~Shenker, G.~W.~Moore and P.~J.~Steinhardt,
  %``Modular cosmology,''
  Phys.\ Rev.\  D {\bf 52}, 3548 (1995)
  [arXiv:hep-th/9503114].
  %%CITATION = PHRVA,D52,3548;%%

\bibitem{KModInfl}
J.~J.~Blanco-Pillado \etal,
 % ``Racetrack inflation,''
  JHEP {\bf 0411} (2004) 063
  [arXiv:hep-th/0406230];
J.~J.~Blanco-Pillado \etal,
%  ``Inflating in a better racetrack,''
  JHEP {\bf 0609} (2006) 002
  [arXiv:hep-th/0603129].
  %%CITATION = JHEPA,0411,063;%%

\bibitem{kahler}
J.~P.~Conlon and F.~Quevedo,
%  ``Kaehler moduli inflation,''
  JHEP {\bf 0601} (2006) 146
  [arXiv:hep-th/0509012];
  %%CITATION = JHEPA,0601,146;%%
J.~R.~Bond \etal,
%, L.~Kofman, S.~Prokushkin and P.~M.~Vaudrevange,
%  ``Roulette inflation with Kaehler moduli and their axions,''
  Phys.\ Rev.\  D {\bf 75} (2007) 123511
  [arXiv:hep-th/0612197].
  %%CITATION = PHRVA,D75,123511;%%

\bibitem{Cicoli3}
  M.~Cicoli, C.~P.~Burgess and F.~Quevedo,
%  ``Fibre Inflation: Observable Gravity Waves from IIB String
%  Compactifications,''
  JCAP {\bf 0903} (2009) 013 [arXiv:0808.0691 [hep-th]].
  %%CITATION = JCAPA,0903,013;%%

\bibitem{DBI}
 E.~Silverstein and D.~Tong,
  %``Scalar speed limits and cosmology: Acceleration from D-cceleration,''
  Phys.\ Rev.\  D {\bf 70}, 103505 (2004)
  [arXiv:hep-th/0310221];
  %%CITATION = PHRVA,D70,103505;%%
M.~Alishahiha \etal,
  %``DBI in the sky,''
  Phys.\ Rev.\  D {\bf 70}, 123505 (2004)
  [arXiv:hep-th/0404084];
  %%CITATION = PHRVA,D70,123505;%%
K.~Becker \etal,
  %``M-theory inflation from multi M5-brane dynamics,''
  Nucl.\ Phys.\  B {\bf 715}, 349 (2005)
  [arXiv:hep-th/0501130].
  %%CITATION = NUPHA,B715,349;%%

\bibitem{kklt}
S.~Kachru \etal,
  %``De Sitter vacua in string theory,''
  Phys.\ Rev.\  D {\bf 68}, 046005 (2003)
  [arXiv:hep-th/0301240].
  %%CITATION = PHRVA,D68,046005;%%

\bibitem{baumann}
D.~Baumann \etal,
%, A.~Dymarsky, I.~R.~Klebanov and L.~McAllister,
 % ``Towards an Explicit Model of D-brane Inflation,''
  JCAP {\bf 0801} (2008) 024
  [arXiv:0706.0360 [hep-th]].
  %%CITATION = JCAPA,0801,024;%%

\bibitem{Nflation}
S.~Dimopoulos \etal,
%, S.~Kachru, J.~McGreevy and J.~G.~Wacker,
  %``N-flation,''
  JCAP {\bf 0808}, 003 (2008)
  [arXiv:hep-th/0507205].
  %%CITATION = JCAPA,0808,003;%%
  
\bibitem{Grimm}
  T.~W.~Grimm,
  %``Axion inflation in type II string theory,''
  Phys.\ Rev.\  D {\bf 77} (2008) 126007
  [arXiv:0710.3883 [hep-th]].

\bibitem{SW}
E.~Silverstein and A.~Westphal,
  %``Monodromy in the CMB: Gravity Waves and String Inflation,''
  Phys.\ Rev.\  D {\bf 78}, 106003 (2008)
  [arXiv:0803.3085 [hep-th]];
  %%CITATION = PHRVA,D78,106003;%%
L.~McAllister \etal,
  %``Gravity Waves and Linear Inflation from Axion Monodromy,''
  Phys.\ Rev.\  D {\bf 82} (2010) 046003
  [arXiv:0808.0706 [hep-th]].
  %%CITATION = PHRVA,D82,046003;%%

\bibitem{nemanja}
N.~Kaloper, A.~Lawrence and L.~Sorbo,
%  ``An Ignoble Approach to Large Field Inflation,''
  JCAP {\bf 1103} (2011) 023
  [arXiv:1101.0026 [hep-th]].
  %%CITATION = JCAPA,1103,023;%%

\bibitem{cklq}
 J.~P.~Conlon \etal,
  %``Volume Modulus Inflation and the Gravitino Mass Problem,''
  JCAP {\bf 0809} (2008) 011 [arXiv:0806.0809 [hep-th]].
  %%CITATION = JCAPA,0809,011;%%

\bibitem{OtherClosedStringInfl}
A.~Avgoustidis \etal,
%, D.~Cremades and F.~Quevedo,
%  ``Wilson line inflation,''
  Gen.\ Rel.\ Grav.\  {\bf 39} (2007) 1203 [arXiv:hep-th/0606031];
  %%CITATION = GRGVA,39,1203;%%
M.~Badziak and M.~Olechowski,
  %``Volume modulus inflation and a low scale of SUSY breaking,''
  JCAP {\bf 0807}, 021 (2008)
  [arXiv:0802.1014 [hep-th]];
  %%CITATION = JCAPA,0807,021;%%
 H.~X.~Yang and H.~L.~Ma,
  %``Two-field K\'ahler moduli inflation on large volume moduli stabilization,''
  JCAP {\bf 0808} (2008) 024 [arXiv:0804.3653 [hep-th]].
  %%CITATION = JCAPA,0808,024;%%

\bibitem{gkp}
 S.~B.~Giddings \etal,
  %``Hierarchies from fluxes in string compactifications,''
  Phys.\ Rev.\  D {\bf 66} (2002) 106006
  [arXiv:hep-th/0105097].
  %%CITATION = PHRVA,D66,106006;%%

\bibitem{bbhl}
K.~Becker, M.~Becker, M.~Haack, and J.~Louis,
%``Supersymmetry breaking and alpha'-corrections to flux induced potentials,''
JHEP {\bf 0206} (2002) 060 [arXiv:hep-th/0204254].
  %%CITATION = JHEPA,0206,060;%%

\bibitem{bhk}
M.~Berg, M.~Haack, and B.~Kors,
%``String loop corrections to Kaehler potentials in orientifolds,''
JHEP {\bf 0511} (2005) 030 [arXiv:hep-th/0508043].
%%CITATION = HEP-TH 0508043;%%.

\bibitem{bhp}
M.~Berg, M.~Haack and E.~Pajer,
%``Jumping Through Loops: On Soft Terms from Large Volume Compactifications,''
JHEP {\bf 0709} (2007) 031 [arXiv:0704.0737 [hep-th]].
%%CITATION = JHEPA,0709,031;%%

\bibitem{Cicoli1}
  M.~Cicoli, J.~P.~Conlon and F.~Quevedo,
%  ``Systematics of String Loop Corrections in Type IIB Calabi-Yau Flux
%  Compactifications,''
  JHEP {\bf 0801} (2008) 052 [arXiv:0708.1873 [hep-th]].
  %%CITATION = JHEPA,0801,052;%%

\bibitem{cw}
S.~R.~Coleman and E.~Weinberg,
%  ``Radiative Corrections As The Origin Of Spontaneous Symmetry Breaking'',
  Phys.\ Rev.\  D {\bf 7} (1973) 1888;
  %%CITATION = PHRVA,D7,1888;%%
S.~Ferrara \etal,
%, C.~Kounnas and F.~Zwirner,
%  ``Mass formulae and natural hierarchy in string effective supergravities'',
Nucl.\ Phys.\  B {\bf 429} (1994) 589 [Erratum-ibid.\  B {\bf 433} (1995) 255] [arXiv:hep-th/9405188].
  %%CITATION = NUPHA,B429,589;%%

\bibitem{hg}
  G.~von Gersdorff and A.~Hebecker,
  %``Kahler corrections for the volume modulus of flux compactifications,''
  Phys.\ Lett.\  B {\bf 624} (2005) 270
  [arXiv:hep-th/0507131].
  %%CITATION = PHLTA,B624,270;%%

\bibitem{shiftInflation}
  J.~P.~Hsu and R.~Kallosh,
%  ``Volume stabilization and the origin of the inflaton shift symmetry in string theory,''
  JHEP {\bf 0404} (2004) 042
  [arXiv:hep-th/0402047].
  %%CITATION = JHEPA,0404,042;%%

\bibitem{marta}
L.~Covi \etal,
%, M.~Gomez-Reino, C.~Gross, J.~Louis, G.~A.~Palma and C.~A.~Scrucca,
%``de Sitter vacua in no-scale supergravities and Calabi-Yau string models,''
  JHEP {\bf 0806} (2008) 057 [arXiv:0804.1073 [hep-th]].
  %%CITATION = JHEPA,0806,057;%%

\bibitem{martainf}
L.~Covi \etal,
%, M.~Gomez-Reino, C.~Gross, J.~Louis, G.~A.~Palma and C.~A.~Scrucca,
%``Constraints on modular inflation in supergravity and string theory,''
  JHEP {\bf 0808} (2008) 055 [arXiv:0805.3290 [hep-th]].
  %%CITATION = JHEPA,0808,055;%%

\bibitem{blumenhagen}
R.~Blumenhagen, S.~Moster and E.~Plauschinn,
%  ``Moduli Stabilisation versus Chirality for MSSM like Type IIB Orientifolds,''
JHEP {\bf 0801} (2008) 058 [arXiv:0711.3389 [hep-th]].
  %%CITATION = JHEPA,0801,058;%%

\bibitem{LVS}
V.~Balasubramanian \etal,
  %``Systematics of moduli stabilisation in Calabi-Yau flux compactifications,''
  JHEP {\bf 0503} (2005) 007  [arXiv:hep-th/0502058].
  %%CITATION = JHEPA,0503,007;%%

\bibitem{Cicoli2}
  M.~Cicoli, J.~P.~Conlon and F.~Quevedo,
%  ``General Analysis of LARGE Volume Scenarios with String Loop Moduli
%  Stabilisation,''
  JHEP {\bf 0810} (2008) 105 [arXiv:0805.1029 [hep-th]].
  %%CITATION = JHEPA,0810,105;%%

\bibitem{LVS2}
J.~P.~Conlon, F.~Quevedo and K.~Suruliz,
  %``Large-volume flux compactifications: Moduli spectrum and D3/D7 soft supersymmetry breaking,''
  JHEP {\bf 0508} (2005) 007 [arXiv:hep-th/0505076].
  %%CITATION = JHEPA,0508,007;%%

\bibitem{bkq}
C.~P. Burgess, R.~Kallosh, and F.~Quevedo,
%``de Sitter string vacua from supersymmetric D-terms,''
JHEP {\bf 0310} (2003) 056 [arXiv:hep-th/0309187].
%%CITATION = HEP-TH 0309187;%%.

\bibitem{cremades}
S.~L.~Parameswaran and A.~Westphal,
%``de Sitter String Vacua from Perturbative Kahler Corrections and Consistent D-terms'',
JHEP {\bf 0610} (2006) 079 [arXiv:hep-th/0602253];
D.~Cremades \etal,
%, M.~P.~Garcia del Moral, F.~Quevedo and K.~Suruliz,
%  ``Moduli stabilisation and de Sitter string vacua from magnetised D7 branes,''
  JHEP {\bf 0705} (2007) 100
  [arXiv:hep-th/0701154];
  %%CITATION = JHEPA,0705,100;%%
S.~Krippendorf and F.~Quevedo,
%  ``Metastable SUSY Breaking, de Sitter Moduli Stabilisation and Kahler Moduli Inflation,''
  JHEP {\bf 0911} (2009) 039
  [arXiv:0901.0683 [hep-th]].
  %%CITATION = JHEPA,0911,039;%%
 M.~Cicoli \etal,
 %, M.~Goodsell, J.~Jaeckel and A.~Ringwald,
  %``Testing String Vacua in the Lab: From a Hidden CMB to Dark Forces in Flux
  %Compactifications,''
  arXiv:1103.3705 [hep-th].
  %%CITATION = ARXIV:1103.3705;%%

\bibitem{Fup}
A.~Saltman and E.~Silverstein,
%``The scaling of the no-scale potential and de Sitter model building,''
JHEP {\bf 0411} (2004) 066 [arXiv:hep-th/0402135];
%%CITATION = HEP-TH 0402135;%%.
A.~Westphal,
%``de Sitter String Vacua from Kahler Uplifting'',
  JHEP {\bf 0703} (2007) 102 [arXiv:hep-th/0611332].
  %%CITATION = JHEPA,0703,102;%%

\bibitem{geq}
C.~Escoda, M.~Gomez-Reino and F.~Quevedo,
%  ``Saltatory de Sitter string vacua,''
  JHEP {\bf 0311} (2003) 065 [arXiv:hep-th/0307160].
  %%CITATION = JHEPA,0311,065;%%

\bibitem{topol}
A.~D.~Linde,
%  ``Monopoles as big as a universe,''
  Phys.\ Lett.\  B {\bf 327} (1994) 208
  [arXiv:astro-ph/9402031];
  %%CITATION = PHLTA,B327,208;%%
A.~Vilenkin,
%  ``Topological inflation,''
  Phys.\ Rev.\ Lett.\  {\bf 72} (1994) 3137
  [arXiv:hep-th/9402085].
  %%CITATION = PRLTA,72,3137;%%

\bibitem{McW}
  R.~Flauger \etal,
%, L.~McAllister, E.~Pajer, A.~Westphal and G.~Xu,
  %``Oscillations in the CMB from Axion Monodromy Inflation,''
  JCAP {\bf 1006}, 009 (2010) [arXiv:0907.2916 [hep-th]].
  %%CITATION = JCAPA,1006,009;%%

\bibitem{NGinAM}
  S.~Hannestad \etal,
%, T.~Haugbolle, P.~R.~Jarnhus and M.~S.~Sloth,
  %``Non-Gaussianity from Axion Monodromy Inflation,''
  JCAP {\bf 1006}, 001 (2010)
  [arXiv:0912.3527 [hep-ph]].
  %%CITATION = JCAPA,1006,001;%%

\bibitem{BHLR}
W. Buchmuller \etal,
%, K. Hamaguchi, O. Lebedev and M. Ratz,
%``Maximal Temperature in Flux Compactifications,''
JCAP {\bf 0501} (2005) 004 [arXiv:hep-th/0411109];
W. Buchmuller \etal,
%, K. Hamaguchi, O.Lebedev and M. Ratz,
%``Dilaton Destabilization at High Temperature,''
Nucl. Phys. {\bf B699} (2004) 292
[arXiv:hep-th/0404168].

\bibitem{Cicoli4}
  L.~Anguelova, V.~Calo and M.~Cicoli,
  %``LARGE Volume String Compactifications at Finite Temperature,''
  JCAP {\bf 0910} (2009) 025  [arXiv:0904.0051 [hep-th]].
  %%CITATION = JCAPA,0910,025;%%

\bibitem{TI}
D.~H.~Lyth and E.~D.~Stewart,
%  ``Cosmology With A Tev Mass GUT Higgs,''
  Phys.\ Rev.\ Lett.\  {\bf 75} (1995) 201
  [arXiv:hep-ph/9502417];
  D.~H.~Lyth and E.~D.~Stewart,
%  ``Thermal Inflation And The Moduli Problem,''
  Phys.\ Rev.\  D {\bf 53} (1996) 1784
  [arXiv:hep-ph/9510204].
  %%CITATION = PHRVA,D53,1784;%%

\bibitem{barreiro}
  T.~Barreiro \etal,
  %, B.~de Carlos, E.~J.~Copeland and N.~J.~Nunes,
  %``Moduli evolution in the presence of thermal corrections,''
  Phys.\ Rev.\  D {\bf 78}, 063502 (2008)
  [arXiv:0712.2394 [hep-ph]].
  %%CITATION = PHRVA,D78,063502;%%

\bibitem{CQ}
  J.~P.~Conlon and F.~Quevedo,
%  ``Astrophysical and Cosmological Implications of Large Volume String Compactifications,''
  JCAP {\bf 0708} (2007) 019 [arXiv:0705.3460 [hep-ph]].
  %%CITATION = JCAPA,0708,019;%

\bibitem{Acharya}
  B.~S.~Acharya \etal,
%  ``Non-thermal Dark Matter and the Moduli Problem in String Frameworks,''
   JHEP {\bf 0806} (2008) 064 [arXiv:0804.0863 [hep-ph]].
  %%CITATION = JHEPA,0806,064;%%

\bibitem{linde}
  R.~Kallosh and A.~Linde,
%  ``Landscape, the scale of SUSY breaking, and inflation,''
  JHEP {\bf 0412} (2004) 004
  [arXiv:hep-th/0411011].
  %%CITATION = JHEPA,0412,004;%%

\bibitem{sequestering}
R.~Blumenhagen \etal,
JHEP {\bf 0909} (2009) 007 [arXiv:0906.3297 [hep-th]].

\bibitem{curvaton}
  D.~H.~Lyth and D.~Wands,
%  ``Generating the curvature perturbation without an inflaton,''
  Phys.\ Lett.\  B {\bf 524} (2002) 5 [arXiv:hep-ph/0110002];
  %%CITATION = PHLTA,B524,5;%%
D.~H.~Lyth,
%  ``Generating the curvature perturbation at the end of inflation,''
  JCAP {\bf 0511} (2005) 006
  [arXiv:astro-ph/0510443];
  %%CITATION = JCAPA,0511,006;%%
 K.~Ichikawa \etal,
%  ``Non-Gaussianity, Spectral Index and Tensor Modes in Mixed Inflaton and Curvaton Models,''
  Phys.\ Rev.\  D {\bf 78} (2008) 023513
  [arXiv:0802.4138 [astro-ph]].
  %%CITATION = PHRVA,D78,023513;%%

\bibitem{ModReheat}
    G.~Dvali \etal,
    %, A.~Gruzinov and M.~Zaldarriaga,
%  ``A new mechanism for generating density perturbations from inflation,''
  Phys.\ Rev.\  D {\bf 69} (2004) 023505
  [arXiv:astro-ph/0303591];
    %%CITATION = PHRVA,D69,023505;%%
   G.~Dvali \etal,
   %, A.~Gruzinov and M.~Zaldarriaga,
%  ``Cosmological perturbations from inhomogeneous reheating, freezeout, and mass domination,''
  Phys.\ Rev.\  D {\bf 69} (2004) 083505
  [arXiv:astro-ph/0305548];
  %%CITATION = PHRVA,D69,083505;%%
  L.~Kofman,
%  ``Probing string theory with modulated cosmological fluctuations,''
  arXiv:astro-ph/0303614.
  %%CITATION = ASTRO-PH/0303614;%%

\bibitem{StringyCurvaton}
C.~P.~Burgess \etal,
%, M.~Cicoli, M.~Gomez-Reino, F.~Quevedo, G.~Tasinato and I.~Zavala,
  %``Non-standard primordial fluctuations and nongaussianity in string inflation,''
  JHEP {\bf 1008} (2010) 045 [arXiv:1005.4840 [hep-th]].
  %%CITATION = JHEPA,1008,045;%%

\bibitem{StringyMR}
C.~P.~Burgess, M.~Cicoli, M.~Gomez-Reino, F.~Quevedo, G.~Tasinato and I.~Zavala,
  %``Non-standard primordial fluctuations and nongaussianity in string inflation,''
  in preparation.
  %%CITATION = JHEPA,1008,045;%%

\end{thebibliography}
\end{document}